# Metallic liquid hydrogen and likely Al$_2$O$_3$ metallic glass

W. J. Nellis

Department of Physics, Harvard University, 17 Oxford Street, Cambridge, MA 02138, USA.

**Abstract.** Dynamic compression has been used to synthesize liquid metallic hydrogen at 140 GPa (1.4 million bar) and experimental data and theory predict Al$_2$O$_3$ might be a metallic glass at ~300 GPa. The mechanism of metallization in both cases is probably a Mott-like transition. The strength of sapphire causes shock dissipation to be split differently in the strong solid and soft fluid. Once the 4.5-eV H-H and Al-O bonds are broken at sufficiently high pressures in liquid H$_2$ and in sapphire (single-crystal Al$_2$O$_3$), electrons are delocalized, which leads to formation of energy bands in fluid H and probably in amorphous Al$_2$O$_3$. The high strength of sapphire causes shock dissipation to be absorbed primarily in entropy up to ~400 GPa, which also causes the 300-K isotherm and Hugoniot to be virtually coincident in this pressure range. Above ~400 GPa shock dissipation must go primarily into temperature, which is observed experimentally as a rapid increase in shock pressure above ~400 GPa. The metallization of glassy Al$_2$O$_3$, if verified, is expected to be general in strong oxide insulators. Implications for Super Earths are discussed.

## 1 Introduction

Dynamic compression achieves extreme pressures, densities, and temperatures, which can be tuned depending on material and time over which pressure is applied. A dynamic compression is defined to be sufficiently fast to be adiabatic. Dynamic compression is also known as shock compression, though compression by a single shock is a particular type of dynamic compression. Experimental life times are typically ~100 ns or less. Novel liquid and amorphous metals can be synthesized at extreme conditions with this technique. In addition, existing results suggest that similar states of matter can be obtained in strong transparent sapphire (single-crystal Al$_2$O$_3$) by both dynamic and static compression, the latter of which is sufficiently slow to be isothermal. Typical static-pressure experimental lifetimes are seconds or more and highest pressures are achieved in a diamond anvil cell (DAC). Materials at extreme conditions in a DAC can be characterized more straightforwardly than materials under dynamic compression.

As examples of unusual materials, amorphous Al$_2$O$_3$ is probably metallic at 300 GPa (3 million bar) and fluid hydrogen is metallic at 140 GPa. Because of disorder and associated strong electron scattering, in these two cases "metallic" means minimum metallic conductivity (MMC). Metallic fluid H was made by reverberating a shock wave in compressible liquid hydrogen contained between two weakly compressible

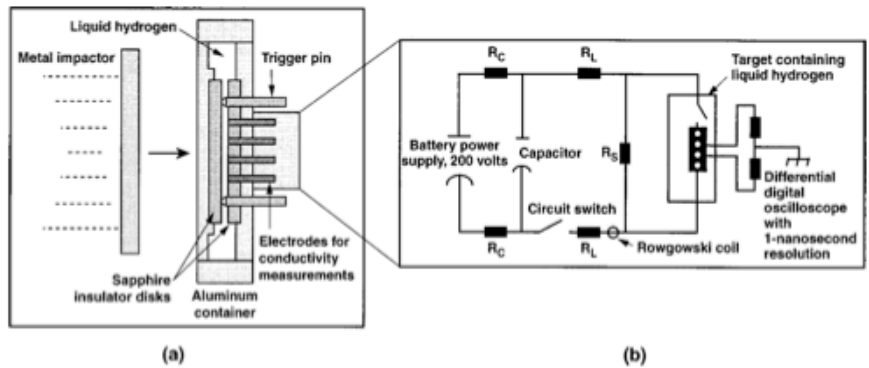

**Fig. 1**. Schematic to measure electrical conductivity of liquid hydrogen at 20 K, contained between sapphire disks, and compressed dynamically from 93 to 180 GPa [1]. Reprinted with permission from W. J. Nellis, S. T. Weir, and A. C. Mitchell, Physical Review B, Vol. 59, Page 3434, (1999). Copyright 1999 by American Physical Society.

*c*-cut sapphire anvils, as illustrated in Figure 1. Shock reverberation (multiple-shock compression) produces substantially higher pressures and densities and lower temperatures than achieved by a single shock. Fluid H reaches minimum metallic conductivity MMC at 140 GPa achieved with impactors accelerated to 5.6 km/s or more with a two-stage light-gas gun [1-3]. Similar results were obtained by Fortov et al using impactors accelerated with high explosives [4]. While metallic fluid hydrogen and amorphous $Al_2O_3$ do not appear to be similar, they are similar in one important way. The strengths of the H-H bond in the diatomic liquid and of the Al-O bond in the three-dimensional strong crystal are essentially equal. For this reason metallization of $Al_2O_3$ is easier to understand once metallization of fluid hydrogen is understood. The purpose of this paper is to discuss the mechanism by which dynamic compression transforms these two very different electrical insulators into poor metals.

## 2 Shock compression

In this discussion shock compression (i) satisfies the Rankine-Hugoniot equations, which are based on conservation of mass, momentum, and energy across the front of a shock wave, (ii) is adiabatic (dQ=0, where Q is energy transported in or out of a system during the compression), and (iii) is one in which entropy S increases during the compression (dS>0) [5]. The rise time of a dynamic-compression wave can be arbitrarily long provided the compression is adiabatic. This definition includes a single wave with the fastest rise time possible from initial to final pressure, density, and internal energy, commonly known as a shock wave. A sound wave is not a dynamic compression because dS=0. However, because only a tiny amount of entropy increase is required for a compression to be dynamic, a sound wave or a purely isentropic compression is the lower bound on a dynamic compression. A dynamic compression is an adiabatic compression intermediate between a shock and an isentropic compression. Dissipation energy is energy used for something other than compression. The free energy F is given by F=U-TS, where U is the internal energy, T is temperature, and TS is dissipation energy.

## 3 Metallization

$H_2$ is one of the most compressible materials and sapphire is one of the least with its high strength (~15 GPa) and 10 eV electronic band gap. H-H and Al-O bond energies are 104 and 107 kJ/mole, respectively, or 4.5 eV per bond for each. Liquid $H_2$ is disordered and has a potential with a well depth of 0.003 eV between pairs of diatomic molecules. The soft pair potential means dynamic compression is rapid and substantial in magnitude, which means thermal equilibrium occurs quickly in hydrogen. In contrast, solid $Al_2O_3$ has a three-dimensional solid structure in which networks of nearest-neighbor Al-O pairs are connected by strong bonds, which implies shock dissipation energy is absorbed primarily by bond breaking and compression is relatively small and slow compared to that of hydrogen. Shock heating of sapphire is probably heterogeneous below ~200 GPa because of its high strength. Fluid H is a poor metal at a shock pressure of 140 GPa [1,2,6]. Remarkably, sapphire is also probably a poor metal at ~300 GPa and probably under both static and shock pressures [7].

Fluid $H_2$ and crystalline $Al_2O_3$ are extreme cases of strength and compressibility, which causes shock dissipation energy to split differently between T and S. In fact, the large strength of sapphire enables recognition of the dominance of S and of T in separate pressure ranges on its shock-compression curve (Hugoniot). At the 100 GPa pressures considered here the transition from strong localized 4.5 eV bonds to itinerant energy bands is driven by dissipation energy TS deposited in shock flows. S is primarily disorder induced by dissociation of $H_2$ and by breaking and bending of Al-O bonds. Simple band closure is insufficient to drive metallization in these two materials; entropy generation is essential. Nevertheless, both materials probably undergo Mott-like transitions to MMC, as discussed below.

## 4 Hydrogen

$H_2$ molecules in the liquid interact via a "soft" Lennard-Jones effective pair potential. Since the well depth, 0.003 eV, is small compared to the dissociation energy, 4.5 eV, $H_2$ undergoes substantial dynamic compression and heating before the onset of dissociation. The compression and temperature achieved are relatively large and homogeneous, which induces thermal equilibrium within the narrow shock front. $H_2$ is an electronic insulator because both electrons are localized on the molecule. $H_2$ dissociates when shock pressure and temperature, actually TS, are sufficiently large.

The relative amounts of equilibrated T and S in the dissociated fluid are determined by the relative amounts of dissociation (S) and T that minimize the free energy. Electrons from atoms produced by dissociation are mobile and form an energy band. At sufficiently high density a system with one electron per atom is a metal. At lower density and finite temperature it is a semiconductor. Metallic fluid H is formed at 140 GPa, 9-fold compression of liquid $H_2$, and ~3000 K. Because of the high density the Fermi energy is 19 eV and $T/T_F$~0.01 and this fluid is highly degenerate. Because the electron mean free path is about the average distance between atoms, the measured electrical conductivity is 2000/ohm-cm, which is Mott's MMC. Similar results have been measured for liquid $N_2$ and $O_2$ to produces metallic fluid N and O at 100 GPa pressures [6,8].

Hensel et al measured values of electrical conductivities of liquid Cs and Rb at 2000 K in a cylindrical system [9], which are similar to those of H, N, and O. On heating to 2000 K near ambient pressure, fluid Rb and Cs are semi-conducting and must be compressed with ~5 MPa to drive them into a state with MMC. MMCs for H, N, O, Rb, and Cs are all ~2000/(ohm-cm). At densities below that needed for MMC in these semiconducting fluids, the density dependences of the logarithms of the electrical conductivities of all five elemental fluids scale with radial extent of the quantum mechanical electron densities of the respective atoms. H, Rb, and Cs atoms are spherically symmetric. We assumed N and O atoms can be treated as spherically symmetric, based on thermal averaging in the fluid at estimated temperatures of a few 1000 K. Radial extents of the outermost 2p electrons of N and O were calculated in the Hartree-Fock-Slater approximation using integrals evaluated by Herman and Skillman [10].

Mott deduced the condition that insulating crystals at T=0 K undergo a discontinuous increase in carrier concentration at $a_B/D^{-1/3}$=constant, where $a_B$ is the effective Bohr radius of the outermost electron(s), D is number density, $D^{-1/3}$ is the average distance between adjacent atoms, and the constant is difficult to calculate [11]. He estimated that the constant is about 0.3 for Ge. At metallization $D^{1/3}a_B$ = ~0.38 for fluid H, N, O, Cs, and Rb [6]. Because temperatures are finite, these fluids are semiconductors at densities less than required for MMC. The largest $D^{1/3}a_B$ can be is 0.5, which corresponds to coincidence of maxima of charge density distributions on neighboring atoms. Thus at metallization, overlap of electron wave functions on adjacent H atoms is extremely large. The primary effect of temperature is to dissociate molecules, which lowers the pressure required for metallization, relative to that of condensed $H_2$, which is yet to be measured.

This nonmetal (semiconductor) to metal transition in fluid $H_2$, $N_2$, and $O_2$ is a Mott-like transition. That is, electrons localized on diatomic molecules are delocalized by dissociation to a monatomic fluid. Pressure then increases overlap of atomic wave functions to achieve the metallic state [6].

A common question is whether metallic fluid H is a superconductor. The measured electrical resistance of 500 µohm-cm at 140 GPa and 3000 K demonstrates this fluid is not a superconductor. However, since metallic fluid H is yet to be quenched to a metastable solid glass at room temperature, it is not yet possible to say if the quenched glass superconducts or not. However, quenched H metallic glass is expected to have a resistivity similar to that of the metallic fluid because of dominance of strong electron scattering. Several amorphous elements are known to superconduct. Amorphous Bi and Ga become superconducting at Tc=~6.0 K and ~7.5 K, respectively [12]. Crystalline Bi does not superconduct and crystalline Ga superconducts at 1.1 K. Thus, there is no a priori reason why metastable H metallic glass cannot superconduct, nor have a high $T_c$.

Many high-$T_c$ superconducting oxides are Mott insulators in the pure compound. That is, electron correlations localize electrons even though band calculations indicate the material should be a metal. To make Mott insulators metallic, they must be doped with carriers. In a certain range of carrier density, those oxides superconduct at temperatures $T_c$ ranging from 30 to 160 K [13]. In the normal state those oxides have room-temperature electrical resistivities of a few 100 µohm-cm. Liquid $H_2$ is an electrical insulator composed of diatomic molecules with 2 electrons localized on each molecule, analogous to electron localization in a Mott insulator. To make $H_2$ metallic the molecular bond must be broken, which in a sense "dopes" the fluid with electron carriers. The resistivity of H metallic glass is expected to be ~500 µohm-cm, comparable to that of high-$T_c$ oxides. Despite these similarities, the question of superconductivity in H glass can only be answered by experimental measurement of $T_c$ and that remains to be done. Other possible uses of metastable H glass, if it could be quenched, are discussed elsewhere [14].

## 5 Sapphire

At ambient, sapphire is one of the strongest materials known and is a large-band gap insulator. Unfortunately, shock compressed sapphire becomes opaque above 100 to 130 GPa [15]. *c*-cut sapphire was chosen as anvils in Figure 1 because it had traditionally been used as optical windows in shock experiments up to 100 GPa. Unfortunately, it is not yet possible to measure the temperature of shock-compressed metallic hydrogen from its Planck spectrum emitted through a transparent sapphire anvil at 140 GPa. Sapphire was investigated in the elastic-plastic range of shock pressures (16 to 86 GPa) [16] (i) to try to understand the nature of shock-induced defects causing shocked sapphire to become opaque above 100 GPa and (ii) to try to identify a direction of shock propagation in the hexagonal lattice in which sapphire remains the most transparent to the highest shock pressures. When those results were combined with previous experiment and theory, the conclusions were well beyond the imagined outcome prior to those experiments.

Sapphire single crystals were investigated with a Velocity Interferometer for a Surface of any Reflector (VISAR). Crystals were cut so that shock propagation would be in one of seven different crystallographic directions in the hexagonal crystal structure. A planar shock in a crystal was generated by impact of a flat Al plate accelerated to 1.2, 1.8, or 5.2 km/s, which produced stresses in sapphire of 16, 23, and 86 GPa, respectively. The impact launched a sharp planar shock (rise time <1 ns) into a crystal. The VISAR measured the temporal velocity profile that exited the rear surface of the crystal. Ideally the elastic and plastic waves have ~1 ns rise times to steady states. In strong materials elastic rise times are typically ~1 ns and 10 to 20 ns for plastic waves.

For sapphire shocked to 23 GPa, measured rise times of the plastic waves ranged from 80 to 300 ns in six of the seven crystallographic directions investigated. In the seventh direction rise times to steady elastic and plastic waves were 1 ns, typical of a fluid. Fortunately, an enormous body of experimental data and recent theory has been accumulated for $Al_2O_3$, with which to reach an explanation for the long rise times. Basically, sapphire disorders continuously via mechanical failure of its strong corundum lattice up to 100 GPa pressures, followed by incomplete transitions to new phases up to about 400 GPa, by which sapphire is amorphous. Dissipative shock energy below ~400 GPa is absorbed primarily in entropy of disorder [17]. Around 300 GPa sapphire is probably an amorphous metal [7].

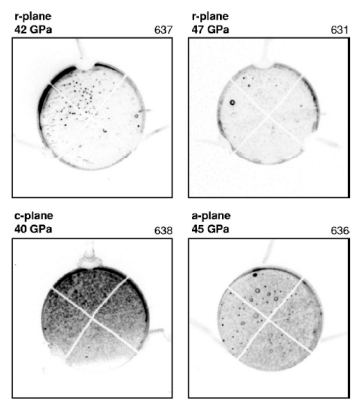

**Fig. 2** Shock-induced emission (dark regions) for *r*-, *c*-, and *a*-cut sapphire crystals at shock pressures indicated [22]. Reprinted with permission from D. E. Hare, N. C. Holmes, D. J. Webb, Physical Review B, Vol. 66, Page 014108, (2002). Copyright 2002 by American Physical Society.

The fact that entropy can be a major mechanism of shock dissipation has essentially been ignored since 1970 when McQueen et al published their tables of physical parameters, including entropy, on Hugoniots of numerous materials [18]. It has generally been assumed since then that temperature is the mechanism of shock dissipation. The fact that entropy dominates dissipation up to 400 GPa on the sapphire Hugoniot is suggested by the fact that Hugoniot data of sapphire up to 340 GPa [19] are virtually coincident with the theoretical Hugoniot [20]. Above ~400 GPa Hugoniot pressure increases dramatically, which implies thermal pressure increases dramatically [21], which implies temperature increases dramatically. The latter suggests that once entropy generation essentially saturates near 400 GPa additional shock dissipation must and does goes into temperature.

Evidence for shock-induced damage in α-phase corundum is given by heterogeneous hot spots observed in fast photographs taken up to 45 GPa for shock waves traveling in $r$, $c$, and $a$ directions [22], as shown in Figure 2. The $r$ direction subtends an angle of $57^0$ with the $c$ axis, which is perpendicular to the basal plane, which contains the $a$ axis. Thermal-emission gray-body spectra up to 86 GPa have small emissivities of ~0.01 [23] indicative of heterogeneous hot spots, consistent with Figure 2. Twins and other shock-induced defects are observed in crystals shocked and recovered from 23 GPa [24].

Wave profiles of shocks traveling in the $m$ and $c$ directions are shown in Figures 3 and 4, respectively. These are the two extreme cases of the seven orientations investigated. The profile for $m$-cut at 23 GPa (1.8 km/s in Figure 3) shows sharp (~1 ns) rise times to steady states for the first (elastic) and second (plastic) shock waves. The $m$ direction is in the basal plane midway between two $a$ axes. These fast rise times of $m$-cut at 23 GPa are typical of fluids and suggest slip planes in the direction of shock propagation. For this reason $m$-cut crystals are the likely optimal anvil/widow in shock experiments. Shock waves in the $m$ direction are of a type expected to drive fast phase transitions.

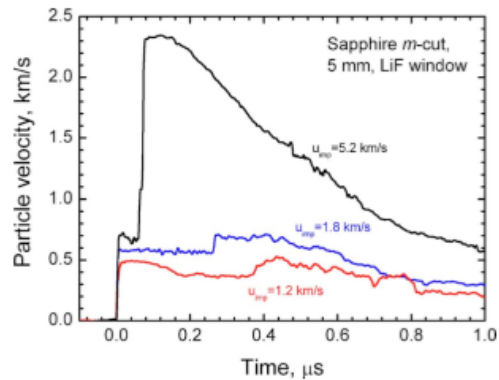

**Fig. 3.** Results of three wave-profile experiments on $m$-cut sapphire crystals 5 mm thick at shock pressure of 16, 23, and 86 GPa [16]. Reprinted with permission from G. I. Kanel, W. J. Nellis, A. S. Savinykh, S. V. Razorenov, A. M. Rajendran, Journal of Applied Physics, Vol. 106, Page 043524, (2009). Copyright 2009 by American Institute of Physics.

The profile for $c$-cut at 23 GPa (1.8 km/s in Figure 4) shows a sharp (~1 ns) rise time for the first (elastic) wave, followed by stress relaxation and then a slow rise time (~300 ns) for the second (plastic) shock wave. A rise time this long has not been reported previously and these profiles scatter substantially in time. The sapphire crystal structure contains Al and O atoms connected by a bond energy of 4.5 eV. Such a material is expected to strongly resist compression because correlated motions of large networks of strongly bonded atoms would have to compress as units, resulting in less compression that takes more time to achieve relative to a fast compression of a liquid in which molecules interact in pairs via an effective Vander Waals potential. The strong Al-O interactions are the probable cause of the long rise times of the plastic wave front. The main effect on the strong crystal is substantial damage at the expense of shock heating. Energy so absorbed mechanically is not available to heat the shocked sapphire, nor to achieve thermal equilibrium. Singly shock sapphire is probably not in thermal equilibrium in the uniform bulk until shock pressure exceeds ~200 GPa or more. Mechanical damage generates substantial entropy. Shock waves in the $c$ direction are expected to drive irreversible phase transitions to an amorphous state.

In a laser-heated diamond anvil cell (DAC) corundum transforms to $Rh_2O_3(II)$-type at 103 GPa and to $CaIrO_3$-type from 130 to 200 GPa [25]. In those cases, compression of corundum at 300 K produced only disordered corundum at high pressures. High-pressure phases were synthesized only by laser heating disordered corundum at high pressures, followed by thermal quenching at high pressures. X-ray spectra of all samples at all pressures consisted of broad individual diffraction peaks, indicative of disordered structures with short-range order. The X-ray spectra of sample at 150 GPa is shown in Figure 5. Ono et al report lattice parameters up to 180 GPa. At 200 GPa only the structure is reported without lattice parameters, presumably because the sample is too disordered. In addition, it is not clear if thermal quenching creates a poorly ordered ground state or quenches a high-temperature phase. Those samples in a DAC contained substantial entropy at 300 K. The lifetimes of shock compression experiments are $\sim 10^8$ times faster than those DAC experiments. The fit to the measured Hugoniot data of sapphire shows no indications of phase transitions up to 340 GPa. Sapphire phase transitions on the Hugoniot are probably frustrated with a continuous evolution of short-range order toward, but never reaching, the long-range order of the equilibrium $CaIrO_3$-type phase. The evolution of short-range order with increasing shock pressure has an associated entropy generation.

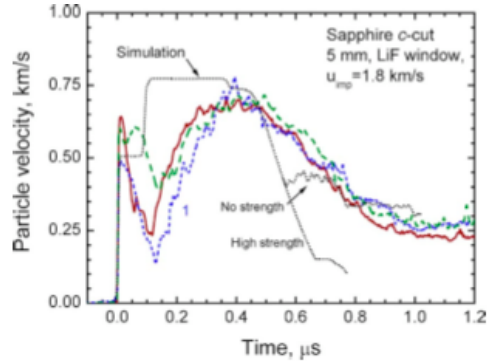

**Fig. 4**. Results of three wave-profile experiments on *c*-cut sapphire crystals 5 mm thick at shock pressure of 23 GPa. Grey curve is result calculated with simple model of ideal elastic-plastic flow for sapphire [16]. Reprinted with permission from G. I. Kanel, W. J. Nellis, A. S. Savinykh, S. V. Razorenov, A. M. Rajendran, Journal of Applied Physics, Vol. 106, Page 043524, (2009). Copyright 2009 by American Institute of Physics.

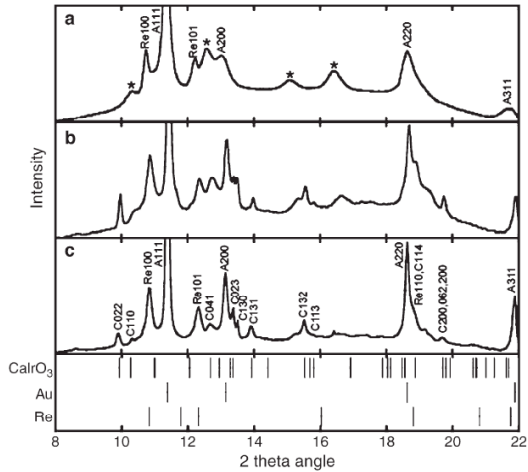

Fig. 5. X-ray diffraction patterns of $CaIrO_3$-type $Al_2O_3$ at 141 GPa: a, corundum phase at 300 K; b, during laser heating at about 2000 K; c, $CaIrO_3$-type at 300K after thermal quench of b. Labels of peaks: C, $CaIrO_3$-type $Al_2O_3$; A, Au pressure gauge; Re, gasket. Stars denote broad peaks of compressed starting material. Vertical bars denote positions of diffraction lines [25]. Reprinted from Earth and Planetary Science Letters, Vol. 246, S. Ono, A. R. Oganov, T. Koyama, H. Shimizu, Stability and compressibility of the high-pressure phases of $Al_2O_3$ up to 200 GPa: Implications for the electrical conductivity of the base of the lower mantle, Pages 326-335, Copyright (2006), with permission from Elsevier.

The question now becomes whether amorphous $Al_2O_3$ is a metal under static and shock compression. The electrical conductivity of sapphire has been measured under shock pressures from 80 to 220 GPa [26]. Extrapolation of the measured conductivities at the two highest shock pressures, 180 and 220 GPa, gives MMC at ~300 GPa (7), the same MMC as for metallic fluid H. How can this be? The likely explanation is disorder under both static and shock compression.

Shock dissipation is dominated by entropy production at these very high pressures. While shock temperatures of sapphire at 100 GPa pressures are yet to be measured, entropy is expected to dominate and shock heating is heterogeneous up to some pressure yet to be determined experimentally. Since directional lattice bonds break at 100 GPa pressures, a reasonable assumption, based on spatial averaging in a glass, is that electron wave functions of Al and O atoms can be treated as spherically symmetric, as for H, N, and O from dissociated $H_2$, $N_2$, and $O_2$ (6). At 300 GPa overlap of Al and O wave functions is substantial and so it is possible for Al and O wave functions to hybridize into an energy band typical of a metal [7]. Because of disorder of the amorphous state, electron scattering is strong because a conduction electron is scattered at every nearest neighbor atom. Strong scattering is not very sensitive to the atom doing the scattering and so MMC is weakly dependent on material. As in H, N, and O the mechanism for conduction in amorphous Al-O is probably a Mott-like transition – metallization is caused by delocalization of electrons initially in bonds whose wave functions hybridize to form metallic bands.

## 6 Generality of High-Pressure Effects in Sapphire

Hugoniots of many strong materials [27], including sapphire [19], $CaTiO_3$ [28], $Gd_3Ga_5O_{12}$ [29], and $CaF_2$ [30], have a characteristic shape. At relatively low pressures $P(\eta)$, where P is pressure, $\eta$ is compression, and $\rho$ and $\rho_0$ are shock and initial density, respectively, has a relatively small slope $(dP/d\eta)$, which increases substantially near 100 GPa for $CaTiO_3$, GGG, and $CaF_2$. The sapphire Hugoniot has a relatively small slope $(dP/d\eta)$ up to 400 GPa, which increases significantly at higher pressures. Thus, the characteristic shapes are similar for all four materials. The dramatic increase in $(dP/d\eta)$ probably occurs when entropy generation saturates, which is determined primarily by material strength. After saturation of entropy, temperature and thus thermal pressure increase dramatically. Sapphire phenomena discussed above are expected to be general in strong electrical insulators.

## 7 Implications for magnetic fields of Super Earths

Planetary magnetic fields are generated by convective motion of conducting fluids. $Al_2O_3$ is a representative planetary oxide. The highest pressure on oxides in Earth is about 130 GPa at about 3000 K at the core-mantle boundary. At these conditions electrical conductivities and viscosities of solid oxides are too small and large, respectively, to produce a significant contribution to Earth's magnetic field. Super Earths are Earth-like planets with masses ~3 to 10 times that of Earth. Oxides in Super Earths reach interior pressures and temperatures much larger than those in Earth and thus the metallization pressure of $Al_2O_3$, ~300 GPa, is expected to be achieved in oxides in some Super Earths. Since $Al_2O_3$ is estimated to melt on the Hugoniot at ~400 GPa [21], which is virtually coincident with the theoretical 300-K isotherm at this pressure, viscosity is expected to decrease near this pressure, perhaps sufficiently to enable convection. Thus, planetary magnetic fields might be produced in some Super Earths, even without Fe cores. However, magnetic fields in such super Earths are expected to be relatively small compared to the surface field of Earth because they would be produced at greater depth below the planets' surface than for Earth.